\documentclass[12pt,fleqn]{article}

\usepackage{graphicx}
\usepackage{dcolumn}
\usepackage{bm}
\usepackage{amsmath}
\usepackage{amssymb}  
\usepackage{subfigure}
\usepackage{psfig,float}
\def\OMIT#1{} 
\begin{document} 

\title{Universal collective fluctuations in gene expression dynamics from yeast to human}

\author{J.C. Nacher{\footnote{These authors contributed equally to this work. \newline
Corresponding authors: nacher@kuicr.kyoto-u.ac.jp, ochiai@kuicr.kyoto-u.ac.jp
}}, T. Ochiai{\footnotemark[1]}, T. Akutsu}     
 
\maketitle
 
\begin{center}
{\it Bioinformatics Center, Institute for Chemical Research, Kyoto University, }\end{center}
\begin{center}
{\it Uji, 611-0011, Japan}
\end{center}

\begin{center}
\end{center} 

\begin{center}

\end{center}

\begin{abstract}
{\small{
In this work, the dynamics of fluctuations in gene expression time series is investigated.  By 
using collected data of gene expression from yeast and human organisms, we found that 
the fluctuations of gene expression level and its average value over time are strongly 
correlated and obey a scaling law.  As this feature is found in yeast and human organisms, it suggests 
that probably this coupling is a common dynamical organizing property of all living systems. To understand 
these observations, we propose a stochastic model which can explain these collective fluctuations, and 
predict the scaling exponent. Interestingly, our results indicate that the observed scaling law emerges 
from the self-similarity symmetry embedded in gene
expression fluctuations.
}}
\end{abstract}


\section{Introduction}

Latest advances and discoveries in network science \cite{doro,bara} has made it possible to obtain a huge 
understanding of the topology of biological complex systems. However, the acquired knowledge 
is restricted by time, as far as the studies focussed mainly on topological-{\it spatial} properties 
of the networks. More recently, some studies have dealt with this issue by unifying 
the study of spatial- and time-dependent observables in networks (systems) in order 
to simultaneously investigate the collective 
behaviour of tens thousand nodes (elements of the systems). In particular, some works 
studied the fluctuations dynamics in natural and technological transport networks \cite{m1,m2}, and described 
the relationship among the flux fluctuations and the average of flux as a scaling-law. On the biological side, experimental 
and theoretical works have also shown that some macroscopic phenomena, as the scale-free distribution, observed 
in gene expression systems (i.e., abundance of mRNA of a gene) emerges
 as a consequence of dynamical and collective self-organizing properties of the system \cite{ueda,ochi}.

In our post-genomic era, DNA microarrays, also known as BioChips and GeneChips, have become 
the latest and most efficient method to simultaneously analyze thousands of genes \cite{chip1,chip2}. By using these 
technologies, the amount of mRNA contained in cells from different tissues and organisms 
can be collected at different time points. Moreover, the abundance of mRNA of a 
gene (i.e., gene expression) fluctuates in time, and it reveals correlations 
among many genes. Although there are currently many efforts to understand the gene regulation 
and transcriptional control, we have only a limited knowledge of these processes due to 
the huge quantity of genes involved and their intrinsic complexity.  In order to obtain a better 
description on the collective behaviour of thousands of genes, we search for dynamical 
organizing principles in gene expression fluctuations, which are common for 
all living organisms. In particular, we analyse the coupling between the 
average on time of gene expression value $m$ and fluctuation  $\sigma$
(i.e., standard deviation) of individual genes, from yeast \cite{cho1} and human \cite{cho2} organisms. Our results 
indicate 
that these magnitudes of gene expression system are related each other 
and depend on a scaling law $\sigma\propto m^\alpha$ with exponent one $\alpha=1$. Intriguingly, the same scaling-law was 
found in \cite{m1} for natural and transport systems as rivers, WWW or highways.

 To explain this observation found in gene expression systems, we developed a 
 theoretical model based on stochastic processes. The general strategy is
  as follows. First, we classified the genes from yeast \cite{cho1} and human \cite{cho2} organisms by 
  the average value of expression level $m$. All the genes which expression level fluctuates 
  around a given $m$ value will belong to the same
  group. Simultaneously, we assume that the gene expression dynamics 
  of each group obeys the Markov property \cite{ochi,kampen}. Secondly, by using the short time transition 
  probability (properly defined later), we construct the stochastic model for gene expression level. Lastly, we 
  solve the model and obtain the same scaling-law observed experimentally. In addition, the relevant 
  scaling-exponent is predicted successfully. The crucial point is that our stochastic approach reveals the existence of the
  self-similar symmetry embedded in the gene expression fluctuations, and the observed scaling-law emerges from this symmetry \cite{ochi2}.

\section{Gene expression time series data}
We used experimental data from two well-known experiments 
on absolute gene expression from yeast \cite{cho1} and human \cite{cho2} organisms. The first 
experiment analysed cell cycle of the budding yeast {\it S. Cerevisiae}, where around 6220 genes 
were monitored. Data of absolute value of gene expression fluctuations were collected 
at 17 time points every 10 min intervals. Data was obtained from WWW site {\it http://genomics.standford.edu}.

The second experiment identifies 
cell-cycle-regulated transcripts in human cells using high-density oligonucleotide 
arrays. Data are collected every 2 hours for 24 hours, what is equivalent 
to almost 2 full cell cycles. The number of genes monitored was around 
35000, and data for absolute value of gene expression fluctuations was 
obtained from WWW site \\
{\it http://www.salk.edu/docs/labs/chipdata}. Further details 
about the experimental set up can be found in \cite{cho1} and \cite{cho2}, respectively.

In Fig. 1, we show the experimental absolute value of gene expression level (vertical axis) vs. time (horizontal axis) 
of a selected group of genes which belong to human organism \cite{cho2}. We see that the gene expression
value of the selected genes fluctuates around the mean value $m$=6000 and $m$=1000.

\section{Fluctuation observables: average and standard deviation of gene expression} 
By using the yeast and human data on gene expression time series, 
we proceed as follows. First, we calculate the average expression value of each gene of yeast datasets \cite{cho1} and human datasets
\cite{cho2}. This expression reads as:
\begin{equation}
m=\frac{\sum_{t=1}^{T} g_t}{T},
\end{equation}
where $g_t$ means the absolute expression level of gene at time step $t$, and $T$ is the total 
number of time steps. Secondly, we calculate the standard deviation (i.e., dispersion) of each gene by using 
the following expression:						
\begin{equation}
\sigma=\sqrt{\frac{1}{T}\sum_{t=1}^{T}(g_t-m)^2}.
\end{equation}
Therefore, for the yeast organism, we obtain 6220 dataset of coupled pairs ($\sigma$, $m$), and 35120 for the human organism. 

\section{Experimental observations}
By plotting the pairs ($\sigma$, $m$) of yeast and human after some binning of the data, we find that 
the following scaling law between fluctuations and average of expression level over time : 
\begin{equation}
\sigma\propto m^\alpha, 
\end{equation}		
where the scaling exponent is one ($\alpha=1$) .

We show these results in Fig. 2(left) (human) and Fig. 2(right) (yeast). 
Both 
organisms (human and yeast) obeys the same scaling-law $\sigma\propto m^\alpha$ with exponent  $\alpha=1$. The 
scaling-law indicates that genes with higher expression level fluctuate more, and in addition, 
the fluctuation size depends linearly on the mean expression value $m$. Therefore, this 
scaling-law is a macroscopic and collective effect 
of the fluctuations shown in Fig. 1

It is worth noticing 
that recent works have reported about this type of scaling law between 
the average of flux and fluctuations in non-biological systems \cite{m1,m2}. In particular, similar 
observations were found in natural transportation systems as river network, and artificial ones 
as WWW and highways \cite{m1}. Intriguingly, these systems show 
the same scaling exponent ($\alpha=1$) that we have also found for gene expression fluctuation 
in yeast and human organisms. In contrast, more 
technological systems as Internet routers and Microchips, revealed a scaling law with exponent 1/2.



In order to understand the origin of this self-organizing principle of gene dynamical systems, we have developed 
a model based on a stochastic approach. This model reveals that the self-similarity symmetry (i.e., surprisingly, 
the gene expression short-time fluctuations 
contain a repeating pattern of smaller and smaller parts that are like the whole, 
but different in size)
in the short-time transition probability reported in our companion 
paper \cite{ochi2}, re-builds the observed scaling law $\sigma\propto m^\alpha$ with exponent $\alpha=1$. We 
will explain the model in detail in next section. 

Finally, it important to remark that while the scaling exponent 
($\alpha=1$) seems to be universal, since as we will explain later, the self-similarity symmetry strongly constrains 
the possible exponents to only $\alpha=1$, the universal nature of the exponent ($\alpha=1/2$) 
is not clear enough as it is also discussed in \cite{m1}.

\section{The model }
\subsection{Stochastic process}
A stochastic process is one in which only the probability distribution 
for future states can be specified, even given the exact knowledge of the present state. Although 
there are many complex non-biological and biological systems which are not stochastic, for our 
gene expression problem a probabilistic approach seems 
plausible and natural. For example, the number of molecules which are involved in 
signal transduction pathways fluctuates from $10^2 - 10^4$ and concurrently, the physical volumes of cells 
are small. Furthermore, an additional source of randomness comes from instrumental noise, 
which exceeds 30$\%$ from chip to chip (GeneChips arrays).

Moreover, experimental and theoretical studies on stochastic noise and fluctuations of gene expression in cells
have recently been carried out  by revealing an inherent ({\it intrinsic noise}) and external 
({\it extrinsic noise}) stochasticity \cite{kuz,elo}, and by providing a 
huge knowledge of the mechanisms involved in subcellular processes.
\cite{paulsson, sato, blake, hasty}.

\subsection{Stochastic differential equation}
Let $\{X(t), 0 \le t <\infty \}$ be a stochastic process. For $(s>t)$ , the conditional probability 
density function $p(y,s|x,t)=p(X(s)=y | X(t)=x )$ is defined as usual manner. For the matter of 
convenience, we often write  $p(y,s)$ for $p(y,s|x,t)$ .

We make use of Ito Stochastic process in order to construct our model. First, we assume the following 
Stochastic Differential Equation (SDE) (see also \cite{ochi2, wong, black} for further details):
 
\begin{eqnarray}\label{eqn: SPDE}
dX^m(t)=\alpha^m(X^m(t)) dt + \beta^m(X^m(t))dW(t),
\end{eqnarray}
where stochastic variable $X^m(t)$ denotes the gene expression level of genes fluctuating around mean value $m$, $\alpha^m(x)$ denotes the drift (i.e. the average of the instantaneous transition of the expression level per unit of time) defined by
\begin{eqnarray}\label{eqn: initial condition}
\alpha^m(x)=\lim_{\epsilon \to 0}\frac{1}{\epsilon}\int (y-x) T_\epsilon^m(y,x) dy ,
\end{eqnarray}                               
$\beta^m(x)$ denotes the diffusion (i.e., the variance ofthe instantaneous transition the gene expression level per unit of time) defined by
\begin{eqnarray} 
\beta^m(x)=\Bigl(\lim_{\epsilon \to 0}\frac{1}{\epsilon}\int (y-x)^2 T_\epsilon^m(y,x) dy\Bigr)^{\frac{1}{2}},
\end{eqnarray}                             
and  $W(t)$ denotes the Wiener process \cite{kampen}. Here, $T_{\epsilon}^m(y,x)$  is an Instantaneous Transition 
Probability (ITP) defined by $T_{\epsilon}^m(y,x) = p^m(y,t+\epsilon|x, t)$  for sufficient small $\epsilon$. Details 
of the proof can be found in \cite{kampen}.

\subsection{Experimental Input Data}  
>From the data of $T_{\epsilon}^m(y,x)$,  we can obtain $\alpha^m(x)$  and $\beta^m(x)$. We evaluate 
these two magnitudes by using experimental data \cite{cho1,cho2} and Eqs. (5-6), and we observed 
that behaviour of $\alpha^m(x)$ and $\beta^m(x)$ can be expressed by using the following functions: 
\begin{eqnarray}\label{eqn: initial condition}
\alpha^m(x)=\mu(m-x),
\end{eqnarray}   
\begin{eqnarray}\label{eqn: initial condition}
\beta^m(x)=m((x/m-1)^2+b).
\end{eqnarray}                           
Readers interested in details of derivation of Eqs. (7-8) are referred 
to our companion manuscript \cite{ochi2}. These results obtained by 
using experimental data are interesting since 
expression level of genes follows the same 
tendency independently of the scale of the 
gene expression value (i.e., exhibiting a self-similar dynamics). In addition, same feature 
has been found in a simple organism as yeast and in a complex one as human.

 By substituting Eqs. (7-8) into Eq. (4), we obtain:
 .
\begin{eqnarray}\label{eqn: SPDE}
dX^m(t)=\mu(m-X^m(t))dt + m((X^m(t)/m-1)^2+b)dW(t),
\end{eqnarray} 
This equation is scale-invariant, since if we take different $m^{\prime}$, we recover the original Eq. (9) by substituting   
$X^{m^\prime}(t) \rightarrow X^m(t) m^{\prime}/m$.

\subsection{Solution of Stochastic Equation} 
 The solution of Eq. (9) is given by the following equation:
\begin{eqnarray}
\rho^m(x)=\frac{K}{m((x/m-1)^2+b)^2}\exp\Bigl(\frac{\mu}{(x/m-1)^2+b} \Bigr),   
\end{eqnarray}           
where $m$ is the mean value of gene expression for each gene, $b\approx 0.2-0.3$, and $\mu=1$. $\rho^m(x)$ indicates the
probability distribution of genes with expression level $x$ which fluctuates around mean value $m$. In \cite{ochi2}, we 
find that Eq. (10) shows a good agreement with experimental data from yeast \cite{cho1} and human {\cite{cho2} organisms. Furthermore, the 
formula $\rho^{cm}(x)=\frac{1}{c}\rho^m(x/c)$ holds for arbitrary real number $c$.

\subsection{ Fluctuation observables obtained by the model.}
\paragraph{Average value of gene expression} 
By using Eq. (10), we can calculate the average expression value as follows:
\begin{equation}
E[x]=\int x\rho(x)dx=m.
\end{equation} 					

\paragraph{Standard deviation of gene expression}
By using the formula $\rho^{cm}(x)=\frac{1}{c}\rho^m(x/c)$ in previous section, we obtain the variance $V[x]\equiv\sigma^2$ as follows:
\begin{equation}
V[x]=\int (x-m)^2 \rho^m(x)dx \propto m^2.			 
\end{equation}
Then, by using Eqs. (11) and (12), we find the observed scaling law in gene expression systems as: 
\begin{equation}
\sigma\propto m^\alpha,
\end{equation}
with exponent $\alpha=1$.

\section{Conclusions}
In summary, we have analysed datasets of gene expression time series from yeast and human
organisms and we have found that the coupling between the average expression level and fluctuations follows 
a scaling-law with exponent one $\alpha=1$. Therefore, as we have found the 
same scaling-law for organisms as yeast and human, we believe that it 
probably indicates that this is a universal feature of the gene expression dynamics
 in all the living organisms. 

Furthermore, in order to explain this observation in gene expression fluctuations, we have proposed 
a stochastic model, which is able to 
reproduce the observed scaling-law and it also generates 
the relevant scaling exponent. Precisely, we show that this 
scaling law with mysterious exponent one emerges 
from one generic mechanism: self-similarity symmetry \cite{ochi2}.

Interestingly, the same scaling-law and the same exponent $\alpha=1$
were recently found in natural transport systems as rivers and highways \cite{m1}. Although in our model the scaling-law and
the exponent one is re-built from the self-similarity symmetry embedded in the gene expression fluctuations, we believe that 
this self-similar property is not only a feature of gene expression systems and it could also be found in natural transport systems and 
technical and artificial dynamical networks. Therefore, our stochastic approach and this symmetry could explain 
the origin of scaling law of fluctuation dynamics in biological and nonbiological systems in a broader scope.



Moreover, as 
this model has been used successfully to uncover other scaling 
properties of gene expression dynamics \cite{ochi,ochi2} and to analyze the gene correlation dynamics \cite{ochi3}, it indicates that 
it contains valuable information about the gene expression dynamics 
and it is a useful tool for studying it. Therefore, further advances 
towards this direction, and in particular, about the study of 
specific biological processes in gene regulatory architectures 
by using these stochastic concepts and techniques are encouraged.

\newpage

\begin{figure}[htb]
\setlength{\unitlength}{1cm}
\begin{picture}(15,12)(-1,-1)
\put(-1,-1){\includegraphics[scale=0.65]{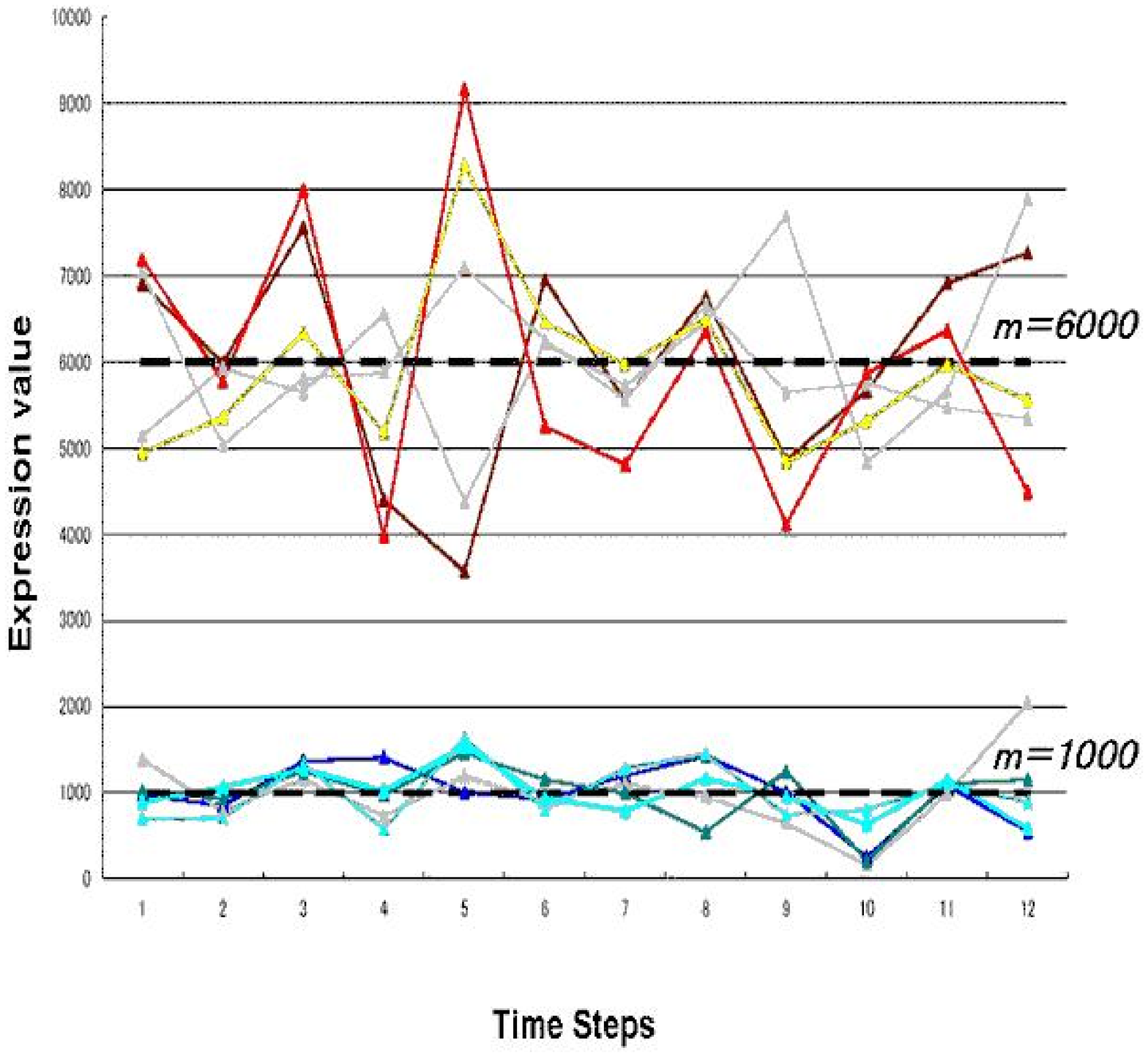}}
\end{picture} 
\caption{\small{We show the experimental absolute value of gene expression level (vertical axis) vs. time (horizontal axis) 
of a selected group of genes which belong to human organism \cite{cho2}. We see that the gene expression
value fluctuates around the mean value $m$=6000 and $m$=1000. By observing this figure, we see that genes with high expression level
fluctuates more (i.e., jump size is large}, which in the end is the origin of the scaling-law $\sigma\propto m^\alpha$ with $\alpha=1$. In Fig. 2, we 
show more clearly this phenomena by using the appropriate observables.}
\label{fig: construction}
\end{figure}

\begin{figure}[htb]
\setlength{\unitlength}{1cm}
\begin{picture}(15,12)(-1,-1)
\put(-2,0){\includegraphics[scale=0.6]{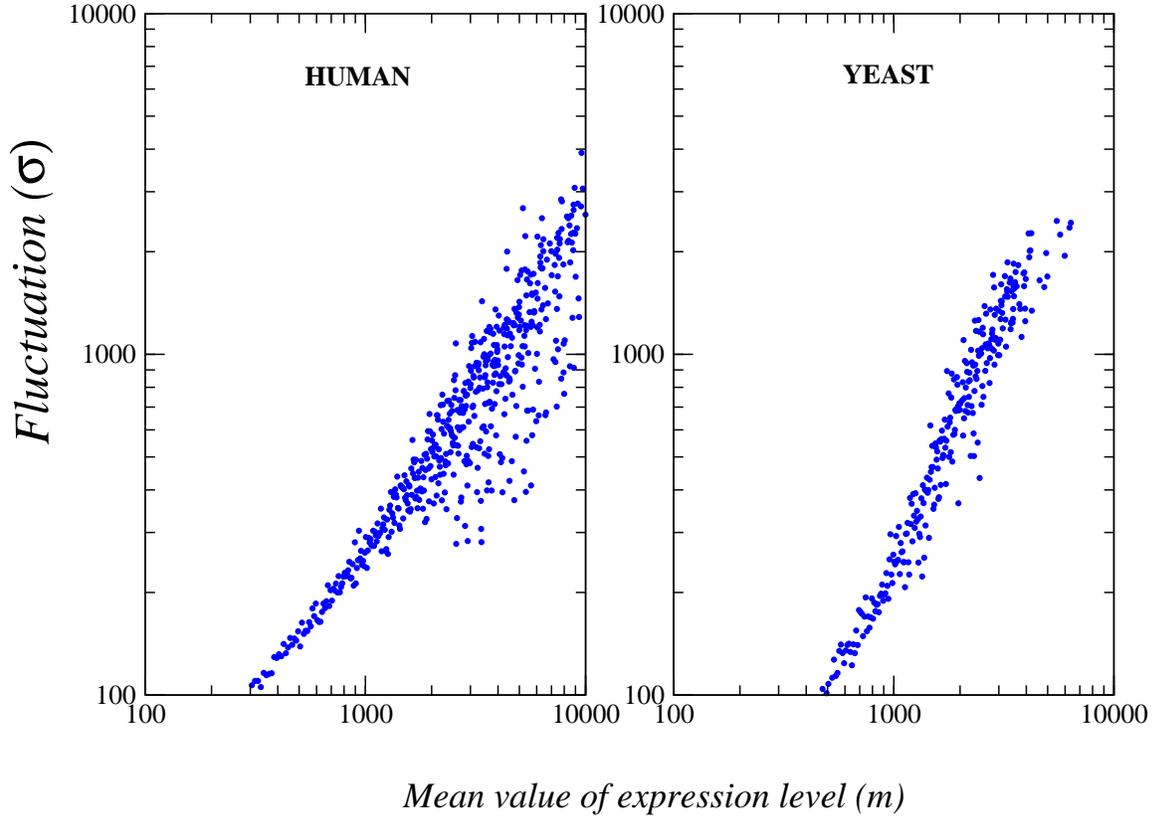}}
\end{picture} 
\caption{\small{
We show the experimental results of the coupling of the fluctuations $\sigma$  
and the average of gene expression level $m$. Horizontal axis denotes 
the average of gene expression level $m$ and vertical axis denotes the 
fluctuations $\sigma$  for each gene expression data. Left) The analyzed data 
corresponds to human organism. Right) The analyzed data corresponds to 
yeast organism. Figures are in log-log scale. We see that both 
organisms (human and yeast) obeys the same scaling-law $\sigma\propto m^\alpha$ with exponent  $\alpha=1$. The 
scaling-law indicates that genes with high expression level fluctuate more, and in addition, 
the fluctuation size (i.e., jump size) depends linearly on the mean value of gene expression $m$. Therefore, this 
scaling-law is a macroscopic and collective effect 
of the fluctuations shown in Fig. 1. It is worth noticing that this scaling-law 
was also found in non-biological systems as for example, rivers, highways and World Wide Web \cite{m1}, with the same exponent $\alpha=1$. 
}}
\label{fig: construction}
\end{figure}


\vspace{0.5cm}
\noindent

\begin{thebibliography}{99}


\bibitem{doro} S.N. Dorogovtsev and J.F.F. Mendes, Evolution of Networks: From Biological Nets 
to the Internet and WWW, (Oxford University Press, Oxford, 2003).

\bibitem{bara} A.-L. Barab\'{a}si and Z.N. Oltvai, Network Biology: Understanding the Cells's 
Functional Organization, Nature Review Genetics {\bf 5}, 101-113 (2004).




\bibitem{m1} M. Argollo de Menezes and A.-L. Barab\'{a}si, Fluctuations in Network Dynamics, Physics Review Letters, {\bf 92}, 028701 (2004).

\bibitem{m2} M. Argollo de Menezes and A.-L. Barab\'{a}si, Separating Internal and External 
Dynamics of Complex Systems, Physics Review Letters, {\bf 93}, 068701 (2004).



\bibitem{ueda} H. Ueda, et al., Universality and flexibility in gene expression from bacteria to human,  Proc. Natl. Acad. Sci. U.S.A, {\bf 101}, (11) 3765 (2004).
\bibitem{ochi} T. Ochiai, J.C. Nacher and T. Akutsu, A constructive approach to gene expression dynamics, Physics Letters A, {\bf 330}, 313-321, (2004).
\bibitem{chip1} P.O. Brown, D. Botstein, Exploring the New World of the genome with DNA microarray, Nature Genetics, {\bf 21}, 33-37 (1999). 
\bibitem{chip2} D.J. Lockhart, et al., Expression monitoring by hybridizidation to high-density oligonucleotide arrays, Nature Biotechnology,
{\bf 14} (13), 1675-80 (1996).
\bibitem{cho1} R.J. Cho, et al. A genome-wide transcriptional analysis of the mitotic cell cycle. Mol. Cell Biol. {\bf 2}, 65-73 (1998).
\bibitem{cho2}	R.J. Cho, et al. Transcriptional regulation and function during the human
 cell cycle. Nature Genetics, {\bf 27}, 48-54 (2001). 
\bibitem{kampen} N.G. Van Kampen, Stochastic processes in Physics and Chemistry, Elsevier, Amsterdam (1992).
\bibitem{ochi2}	T. Ochiai, J.C. Nacher, T. Akutsu, Symmetry and Dynamics in gene expression: Self-similarity symmetry governs the gene
expression dynamics. e-print archive
{\it http://arxiv.org/abs/q-bio.BM/0503003}. {\it Submitted} (2005).


\bibitem{kuz} V.A. Kuznetsov, G.D. Knott, R.F. Bonner, General statistics of stochastic process of gene expression in Eukaryotic cells. Genetics {\bf 161}, 1321 (2002).
\bibitem{elo} M.B. Elowitz, A.J. Levine, E.D. Siggia, P.E. Swain, Stochastic gene expression in a single cell. Science {\bf 297}, 1183 (2002).

\bibitem{paulsson} J. Paulsson, Summing up the noise, Nature, {\bf 427}, 415 (2004).
\bibitem{sato} K. Sato, Y. Ito, T. Yomo, K. Kaneko, On the relation between fluctuation and response in biological systems, Proc. Natl. Acad. Sci. U.S.A {\bf 100}, 14086 (2003).
\bibitem{blake} W.J. Blake, M. Kaern, C.R. Cantor, J.J. Collins, Noise in eukaryotic gene expression, Nature {\bf 422} 633 (2003).
\bibitem{hasty} J. Hasty, J. Pradines, M. Dolnik, J.J. Collins, Noise-based switches and amplifiers for gene expression, Proc. Natl. Acad. Sci. U.S.A  {\bf 97} (5) 1075 (2000).


\bibitem{wong} E. Wong, {\it Stochastic Processes in Information and Dynamical Systems}, Ed. New York, McGraw-Hill (1971).


\bibitem{black} T. Mikosch, {\it Elementary Stochastic Calculus with Finance in View}, World Scientific Publishing Co. Pte. Ltd (1998).

\bibitem{ochi3}	T. Ochiai, J.C. Nacher, T. Akutsu, A stochastic approach to multi-gene expression dynamics. e-print archive
{\it http://arxiv.org/abs/q-bio.BM/0502015}. Physics Letters A, {\it in press} (2005).

\end{thebibliography}
\end{document}